\documentclass[preprint,12pt,number]{elsarticle}
\usepackage{natbib}
\usepackage{amsmath,cases}   
\usepackage{graphicx}  
\usepackage{amssymb}
\usepackage{color,graphics,epsfig}
\journal{Journal of Theoretical Biology}

\begin{document}

\def\be{\begin{equation}}
\def\ee{\end{equation}}
\def\bea{\begin{eqnarray}}
\def\eea{\end{eqnarray}}
\def\bml{\begin{mathletters}}
\def\eml{\end{mathletters}}
\def\l{\label}
\def\b{\bullet}
\def\eqn#1{(~\ref{eq:#1}~)}
\def\no{\nonumber}
\def\av#1{{\langle  #1 \rangle}}
\newcommand{\tc}{\textcolor}
\begin{frontmatter}
\title{Fixation probability of a nonmutator in a large population of asexual mutators}

\author{Kavita Jain and Ananthu James}

\address{Theoretical Sciences Unit,
Jawaharlal Nehru Centre for Advanced Scientific Research, Jakkur,
Bangalore 560064, India} 

\begin{abstract}
In an adapted  population of mutators in which most mutations are deleterious, a nonmutator that lowers the mutation rate is under indirect selection and can sweep to fixation.  Using a multitype branching process, we calculate the fixation probability of a rare nonmutator in a large population of asexual mutators. We show that when beneficial mutations are absent, the fixation probability is a nonmonotonic function of the mutation rate of the mutator: it first increases sublinearly and then decreases exponentially. We also find that beneficial mutations can enhance the fixation probability of a nonmutator. Our analysis  is relevant to an understanding of recent experiments in which a reduction in the mutation rates has been observed. 
\end{abstract}

\end{frontmatter}


\section{Introduction}\label{intr}

Following the conclusion that  mutation rates are subject to the action of evolutionary forces \citep{Sturtevant:1937}, there have been various experimental \citep{Chao:1983,Trobner:1984,Sniegowski:1997,Giraud:2001,Notleymcrobb:2002,Lynch:2010b,Sung:2012,Mcdonald:2012,Wielgoss:2013,Singh:2017} and theoretical \citep{Kimura:1967,Leigh:1973,Taddei:1997,Tenaillon:1999,Johnson:1999a,Palmer:2006,Wylie:2009,Sniegowski:2010,Lynch:2011,Desai:2011,Jain:2013,James:2016,James:2016b,Good:2016} works on the evolution of mutation rates. Many theoretical and empirical studies on adapting populations \citep{Raynes:2014}  have shown that the mutator alleles that elevate the mutation rates can reach a high frequency by generating beneficial mutations and hitchhiking with them \citep{Smith:1974}.

However, once the population has adapted to an environment, due to high rate of production of deleterious mutations, the mutators experience a selective disadvantage and the nonmutator allele that lowers the mutation rate is favored due to indirect selection. Indeed, in a long term evolution experiment on \textit{E. coli}, the frequency of mutators increased in three out of twelve replicate lines while the population was rapidly adapting \citep{Sniegowski:1997}. But when the rate of fitness increase slowed down considerably, one of the mutator lines experienced a decrease in its mutation rate \citep{Wielgoss:2013}. Several other experiments  have also provided evidence for the rise in frequency of nonmutator allele in an adapted population \citep{Trobner:1984,Notleymcrobb:2002,Mcdonald:2012,Singh:2017}.

In this article, we are interested in a theoretical understanding of the evolution of mutation rates in adapted populations. In particular, using a multitype branching process \citep{Harris:1963,Patwa:2008}, we study the fixation probability of a nonmutator allele in a large asexual population of mutators that is moderately well adapted.  In a recent work by us \citep{James:2016}, this question was addressed when the nonmutator arises in the background of strong mutators whose mutation rate is ten to hundred fold higher than the nonmutator \citep{Sniegowski:1997,Oliver:2000}.  However, as experiments show that the mutation rate decreases merely by a factor two to three in an adapted population \citep{Mcdonald:2012,Wielgoss:2013}, here we undertake a more general investigation by allowing the nonmutator's mutation rate to be comparable to that of the mutator. We also address how beneficial mutations in the mutator that increase its fitness affect the nonmutator fixation. Unlike in \citep{James:2016} where this question was studied in a limited parameter regime, here aided by an exact solution for the population frequency distribution that was obtained recently \citep{Jain:2016}, we explore the parameter space completely. 

The article is organised as follows: we define the model, and describe the simulation details and a multitype branching process in Sec.~\ref{mm}.  The results for the fixation probability of a nonmutator are obtained in Sec.~\ref{results}  followed by a discussion in Sec.~\ref{disc}. 

\section{Models and methods}
\label{mm}

\subsection{Individual-based computer simulations}
\label{sto_sim}

We consider an asexual population of mutators in which a mutation, irrespective of its location on the genome, changes fitness by a constant factor. Thus the fitness of an individual carrying $k \geq 0$ deleterious mutations (or, in the $k$th fitness class) is given by 
\be 
\label{fitfunction}
W(k)=(1-s)^{k} ~,~
\ee 
where $0 < s < 1$ is the selection coefficient.  The population is of finite size $N$ and evolves via the standard Wright-Fisher dynamics \citep{Ewens:1979} in which a parent in the fitness class $k$ is selected with a probability equal to $W(k)/N {\overline W(t)}$, where ${\overline W(t)}$ is the average fitness of the population at generation $t$. The selection step is followed by mutations; we employ a single-step mutation model in which mutations are allowed to occur in the neighboring fitness classes only. In an individual carrying $k > 0$ unfavorable mutations, a deleterious mutation occurs at rate $U_d$ to fitness class $k+1$  and a beneficial one at rate $U_b < U_d$ to fitness class $k-1$. In the fittest individual, only deleterious mutations are allowed. 

 Motivated by a long-term evolution experiment on {\it E. coli} in which the nonmutator allele 
emerged in a mutator population when its fitness had almost saturated \citep{Sniegowski:1997,Wielgoss:2013}, we allow the nonmutator to appear after the mutator population has attained a steady state  \citep{James:2016,James:2016b}. 
The invading nonmutator with deleterious and beneficial mutation rates $u_d$ and $u_b$, respectively, carrying $k$ unfavorable mutations arises in the mutator subpopulation in the $k$th fitness class with a probability equal to the stationary fraction of that subpopulation.
 
\begin{figure} 
\begin{center}  
\includegraphics[width=1\textwidth,angle=0,keepaspectratio=true]{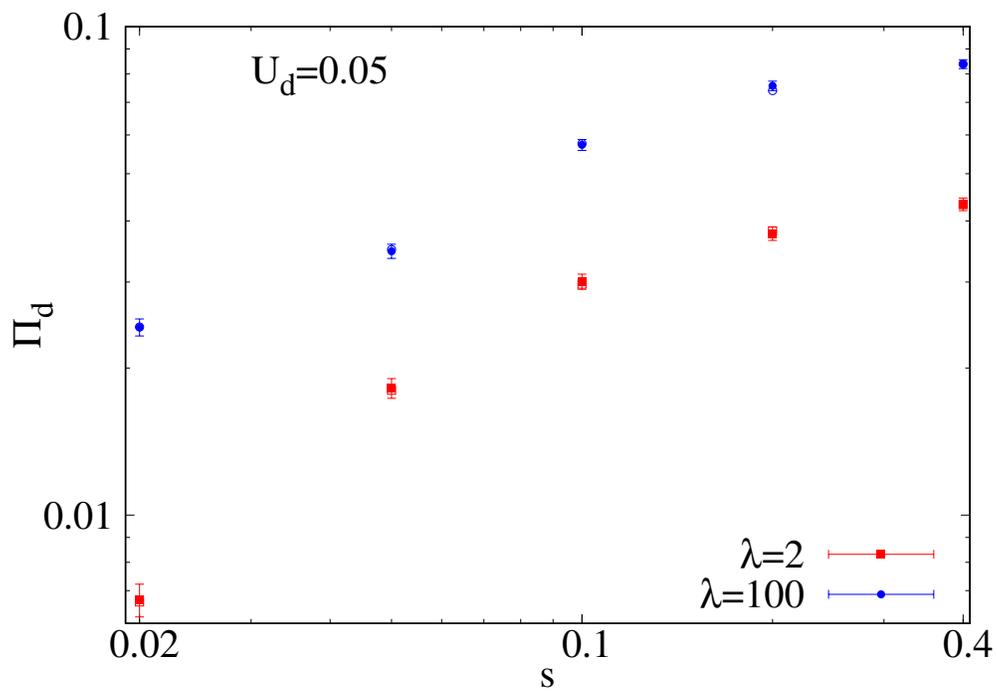}        
 \end{center}
\caption{Variation of the total fixation probability $\Pi_d$ with the selection coefficient  $s$ when beneficial mutations are absent for weak ($\lambda=2$) and strong mutators ($\lambda=100$). The \tc{black}{filled} symbols show the simulation results for population size $N=5000$ (with  error bars representing $\pm 2$ standard error)  and the \tc{black}{open} symbols show the data obtained by numerically solving the recurrence relation (\ref{extprob2}). The overlapping points signify that the agreement between the two methods is very good.}
\label{fig_sdepn}
\end{figure}

We measured the fixation probability of a single copy of nonmutator in a large population of mutators of strength $\lambda > 1$ which is given by the ratio $U_d/u_d=U_b/u_b$. As we are interested in adapted populations in which beneficial mutations are rare, we first ignore the beneficial mutations completely (as discussed in Sec.~\ref{only_del}) and then include beneficial mutations (see Sec.~\ref{comp_mut}). In the former case, as the population size $N$ is finite, Muller's ratchet \citep{Haigh:1978} operates in the mutator population  and there is no true steady state. For this reason, we simulated large enough populations in which the Muller's ratchet clicks very slowly \citep{Jain:2008b} and the mutator population is close to the stationary state of an infinitely large population.   
The fixation probability of a nonmutator was obtained using $10^5$ independent stochastic realizations of the mutator population; the results are shown in Figs.~\ref{fig_sdepn} and~\ref{fig_lambda} when beneficial mutations are ignored and in Fig.~\ref{fig_benmut} when they are taken into account.  
 
\subsection{Multitype branching process}
\label{analys}

In a finite population of mutators, a rare nonmutator allele - although beneficial due to indirect selection - can get lost because of stochastic fluctuations. But if it manages to survive random genetic drift, the nonmutator population can reach a frequency comparable to that of the mutators or even substitute them. Then it is interesting to ask: what is the probability that a rare beneficial allele arising in a large resident population does not go extinct? The branching process \citep{Harris:1963,Patwa:2008} is tailor-made to answer precisely such questions and here we employ it to obtain an analytical understanding of our  simulation results. 
 
Let $1-\pi(k,t)$ denote the extinction probability that a nonmutator arising at generation $t$ in a mutator background with $k$ deleterious mutations is eventually lost. If such a nonmutator gives rise to $n$ offspring in the next generation with probability $\psi_n (k)$, then all the $n$ lineages must go extinct in order to contribute to the probability $1- \pi(k,t)$.  Furthermore, if  mutations are also allowed to occur in the nonmutator from fitness class $k$ to $j$ with a probability $M(k\rightarrow j)$, then summing over the number of offspring produced, we can write \citep{Johnson:2002}
\begin{eqnarray}
\label{extprob1} 
1-\pi(k,t) = \sum_{n=0}^{\infty} \psi_n (k) \left[\sum_{j} M(k\rightarrow j) ~ (1-\pi(j,t+1)) \right]^n  ~.~
\end{eqnarray}
For the Wright-Fisher process described in the last subsection, the offspring number distribution can be approximated by a Poisson distribution ($\psi_n (k)=e^{-\mu_k} \mu_k^n/n!$) with mean $\mu_k$ equal to the relative fitness ${W(k)}/{\overline W(t)}$. We then arrive at 
\begin{eqnarray}
\label{extprob22} 
1-\pi(k,t) =  \exp\left[- \frac{W(k)}{\overline W(t)} \sum_{j} M(k\rightarrow j) ~ \pi(j,t+1) \right]  ~,~
\end{eqnarray}
where we have used that $\sum_j M(k\rightarrow j)=1$. As discussed in Sec.~\ref{sto_sim}, we assume that the nonmutator   appears only after the mutator population has attained a steady state ($t \to \infty$). In this limit, the fixation probability becomes independent of time and in the following, we drop the time argument to denote the stationary state properties.  We are thus required to solve the following nonlinear equation for $\pi(k)$, 
\begin{eqnarray}
\label{extprob2} 
1-\pi(k) =  \exp\left[- \frac{W(k)}{\overline W} \sum_{j} M(k\rightarrow j) ~ \pi(j) \right]  ~.~
\end{eqnarray}

The solutions of the above equation give the fixation probability of a nonmutator in the fitness class $k$. 
But the probability that the nonmutator {\it appears} in the $k$th fitness class is given by the \tc{black}{frequency} $p(k)$ of the mutator population in that fitness class. Thus the total fixation probability obtained by summing over all the mutator backgrounds can be expressed as \citep{Johnson:2002}
\be
\label{fptot}
\Pi = \sum_{k=0}^\infty p(k)~\pi(k) ~.~
\ee  
In (\ref{extprob2}) and (\ref{fptot}) above, we will use the deterministic results for the \tc{black}{equilibrium frequency distribution $p(k)$ and the average fitness ${\overline W}$ of the mutator} as we are working with large mutator populations in which the stochastic fluctuations can be neglected (see Appendix~\ref{app_det} for details). The mutation matrix $M(k \to j)$ is defined in (\ref{M1}) and (\ref{M2}) below. 
The recurrence equation (\ref{extprob2}) along with (\ref{fptot}) are implemented using the software \textit{Wolfram Mathematica}. The numerical results thus obtained are compared with those from stochastic simulations in Figs.~\ref{fig_sdepn} and~\ref{fig_lambda}, and we see a very good agreement. Therefore, in most of the following section, we will discuss results obtained using the multitype branching process.  


\section{Results}
\label{results}

\subsection[short]{Only deleterious mutations}
\label{only_del}

Since the beneficial mutation rates are much smaller than their deleterious counterparts \citep{Perfeito:2007}, as a first approximation, we set the beneficial mutation rates $U_b$ and $u_b$ equal to zero and denote the quantities of interest with a subscript $d$. 

The \tc{black}{frequency distribution of the mutator} $p_d(k)$ is known to be Poisson-distributed  with mean $U_d/s$ \citep{Kimura:1966,Haigh:1978} which gives the mean fitness ${\overline W}_d=\sum_{k=0}^\infty W(k) p_d(k)=e^{-U_d}$ 
(also see Appendix~\ref{app_det}). Furthermore, as for the mutators, we assume that the mutations in the nonmutator occur in the neighboring fitness classes only so that 
\be
M(k\rightarrow k+i)=u_d \delta_{i,1}+(1-u_d) \delta_{i,0} ~,~
\label{M1}
\ee 
with $u_d \ll 1$. From (\ref{extprob2}), we thus obtain
\be
\label{fpk_weak_ss1}
1 - \pi_d(k) = \exp \left[- \frac{(1-s)^k}{e^{-U_d}} \left(u_d \pi_d(k+1)+(1-u_d) \pi_d(k) \right) \right] ~.~
\ee 
Note that $\pi_d(k)=0$ for all $k \geq 0$ is an exact solution of the above equation. 
However, as explained below, a nontrivial solution for the fixation probability exists if the number of mutations carried by the nonmutator are small enough. 

If a nonmutator arising with $k$ detrimental mutations escapes random genetic drift and displaces the resident mutator population, the steady state fitness of the resulting population is given by $e^{-u_d}(1-s)^k$ \citep{Haigh:1978}. The additional factor $(1-s)^k$ reflects the fact that every individual in such a population carries at least $k$ deleterious mutations. Then the maximum number of mutations that the nonmutator can carry 
so as to have a selective advantage over the mutator population is determined by the condition $e^{-u_d}(1-s)^k > e^{-U_d}$ \citep{Johnson:2002}, or $k < k_{\max}$ where 
\be
\label{kmax}
k_{\max} = \lfloor\left(U_d-u_d\right)/s\rfloor ~.~
\ee
Here, $\lfloor x \rfloor$ denotes the maximum integer less than or equal to $x$. Equation (\ref{fpk_weak_ss1}) along with the boundary condition $\pi_d(k_{\max}+1)=0$ can be solved numerically in a straightforward manner for $k \leq k_{\max}$  \citep{Johnson:2002}. 

However, to obtain an analytical solution, we approximate (\ref{fpk_weak_ss1}) using the fact that all the variables ($u_d, U_d, s, \pi_d$) are smaller than one and furthermore, the product $s k < 1$ due to (\ref{kmax}).  
   Taking logarithm on both sides of (\ref{fpk_weak_ss1}) and using the expansion $\ln(1-x)$ $\approx$ $-x-x^2/2$ for $x \ll 1$,  we obtain 
\be
\pi_d(k) +\frac{ {\pi^2_d}(k)}{2} \approx (1+ U_d- s k) \left[u_d \pi_d(k+1)+(1-u_d) \pi_d(k) \right] 
\ee
which can be further simplified to yield  the following nonlinear recursion equation, 
\be
\label{fpk_weak_ws4}
\frac{ {\pi^2_d}(k)}{2} \approx u_d \pi_d(k+1)+(U_d-s k-u_d) \pi_d(k) ~,~ 0 \leq k \leq k_{\max}~.~
\ee
Some remarks are in order: on dividing both sides of the above equation by $(2 s)^2$,  we first note that the scaled fixation probability $\pi_d(k)/(2 s)$ is a function of two scaled mutation rates, {\it viz.}, $u_d/s$ and $U_d/s$. 
The three cases considered in the following subsections are classified according to whether these ratios lie below or above one. (Of course, one can also choose to scale the variables by one of the mutation rates.) Second, for a given integer $k_{\max}$, the model parameters lie in the range $k_{\max} < (U_d-u_d)/s < k_{\max}+1$. Thus, when $k_{\max}$ is large, we may ignore the fact that it is an integer and write $k_{\max} \approx (U_d - u_d)/s$. 
For small integer $k_{\max}$, some simple cases are worked out in Appendix~\ref{app_smallmax}. 

\subsubsection {Case I: $u_d < U_d < s$}
\label{reg1}

In this case, as the selective effect of a deleterious mutation is large, any nonmutator carrying nonzero deleterious mutations gets eliminated from the population and therefore $\pi_d(1)=0$ (this conclusion also follows from (\ref{kmax})). Using this boundary condition in (\ref{fpk_weak_ws4}), we obtain a quadratic equation for $\pi_d(0)$  whose nonzero root  is given by $2(U_d-u_d)$. Furthermore, from (\ref{popfrac_poisson}), since $p_d(0) \approx 1$ when $U_d < s$,  we find that the total fixation probability defined in (\ref{fptot}) is given by  \citep{Lynch:2011,James:2016}
\be
\label{fptot_weak_ss}
\Pi_d = 2(U_d-u_d) ~.
\ee
Thus the total fixation probability is simply twice the fitness advantage $e^{-u_d}-e^{-U_d} \approx U_d-u_d$ conferred by the nonmutator \citep{Haldane:1927b,Lynch:2011,James:2016}.

\subsubsection{Case II: $u_d< s <U_d$}
\label{reg2}

As $U_d$ is the largest variable in this parameter regime, on dividing both sides of (\ref{fpk_weak_ws4}) by $(2 U_d)^2$, we can rewrite it as 
\be
Q^2_d(k)- \left(1-\frac{u_d}{U_d}-\frac{s k}{U_d} \right) Q_d(k) - \frac{u_d}{U_d}Q_d(k+1) = 0 ~,~
\label{scaled}
\ee 
where $Q_d(k)=\pi_d (k)/2 U_d$. 
Since $u_d/U_d$ is the smallest parameter here, we first ignore the terms containing $u_d/U_d$ in the above equation 
and immediately find that the fixation probability decays linearly with the fitness class,  
\be
\label{fpk_weak_ws_sm}
Q_d(k)= 1-\frac{s k}{U_d} ~,~
\ee 
which shows that a nonmutator has a low chance of fixation if it arises in a  mutator background with close to $U_d/s$ mutations. However, as the \tc{black}{mutator frequency} is Poisson-distributed with mean $U_d/s$ (see  (\ref{popfrac_poisson})), the nonmutator is most likely to occur in fitness classes in the neighborhood of $U_d/s$ and therefore such low-fitness classes can still contribute to the total fixation probability. Furthermore, due to the form of (\ref{fpk_weak_ws_sm})  above, the total fixation probability (\ref{fptot}) can be interpreted as the average (positive) deviation from the mean which, by virtue of  (\ref{popfrac_poisson}), is given by $\sqrt{U_d/s}$ and therefore we expect $\Pi_d \propto \sqrt{s U_d}$ (also see (\ref{case2w}) below).

The above discussion applies when the mutator is strong, {\it i.e.}, its mutation rate is much larger than that of the nonmutator ($U_d \gg u_d$) \citep{James:2016}. However, for weak mutators for which $u_d/U_d$ is not negligible,  corrections to the above behavior can be found by expanding the probability $\pi_d(k)$ in a power series in the small  parameter $u_d/U_d$: 
\be
Q_d(k) = 1-\frac{s k}{U_d} + \frac{u_d}{U_d} ~ \Delta(k) ~.
\ee
Substituting the above expression in (\ref{fpk_weak_ws4}) and retaining terms to linear order in $\Delta(k)$, we obtain
\be
Q_d(k) = \frac{\pi_d (k)}{2 U_d}=1-\frac{s k}{U_d}  - \frac{u_d}{U_d} \frac{s}{U_d-s k}  ~,~
\label{case2wk}
\ee
which shows that the nonmutator's chance of fixation is lowered when the mutator is weaker.  

The total fixation probability (\ref{fptot}) obtained by summing over the mutator backgrounds is calculated in Appendix~\ref{app_case2w} and we find that 
\be
\Pi_d = \sqrt{\frac{2 s U_d}{\pi}} \left[1- \frac{u_d}{2 U_d} \ln \left(\frac{{2 s U_d}}{u_d^2} \right) \right] ~.
\label{case2w}
\ee
When the mutation rate $u_d=0$, the second term in the bracket on the RHS vanishes and we recover the result in \citep{James:2016}. The reduction in the fixation probability is, however, not appreciable for moderately strong mutators. For $s=10^{-4}, U_d=10 s$ and $\lambda=50, 100, 200$, using (\ref{extprob2}), we find the fixation probability to be $0.96, 0.98, 0.99$ times the fixation probability when $\lambda \to \infty$, respectively. The corresponding numbers obtained using (\ref{case2w}) are $0.939, 0.962, 0.977$ which overestimate the reduction by $1-2\%$. 

\subsubsection{Case III: $s < u_d < U_d$}
\label{reg3}

\begin{figure} 
\begin{center}  
\includegraphics[width=0.8\textwidth,angle=270,keepaspectratio=true]{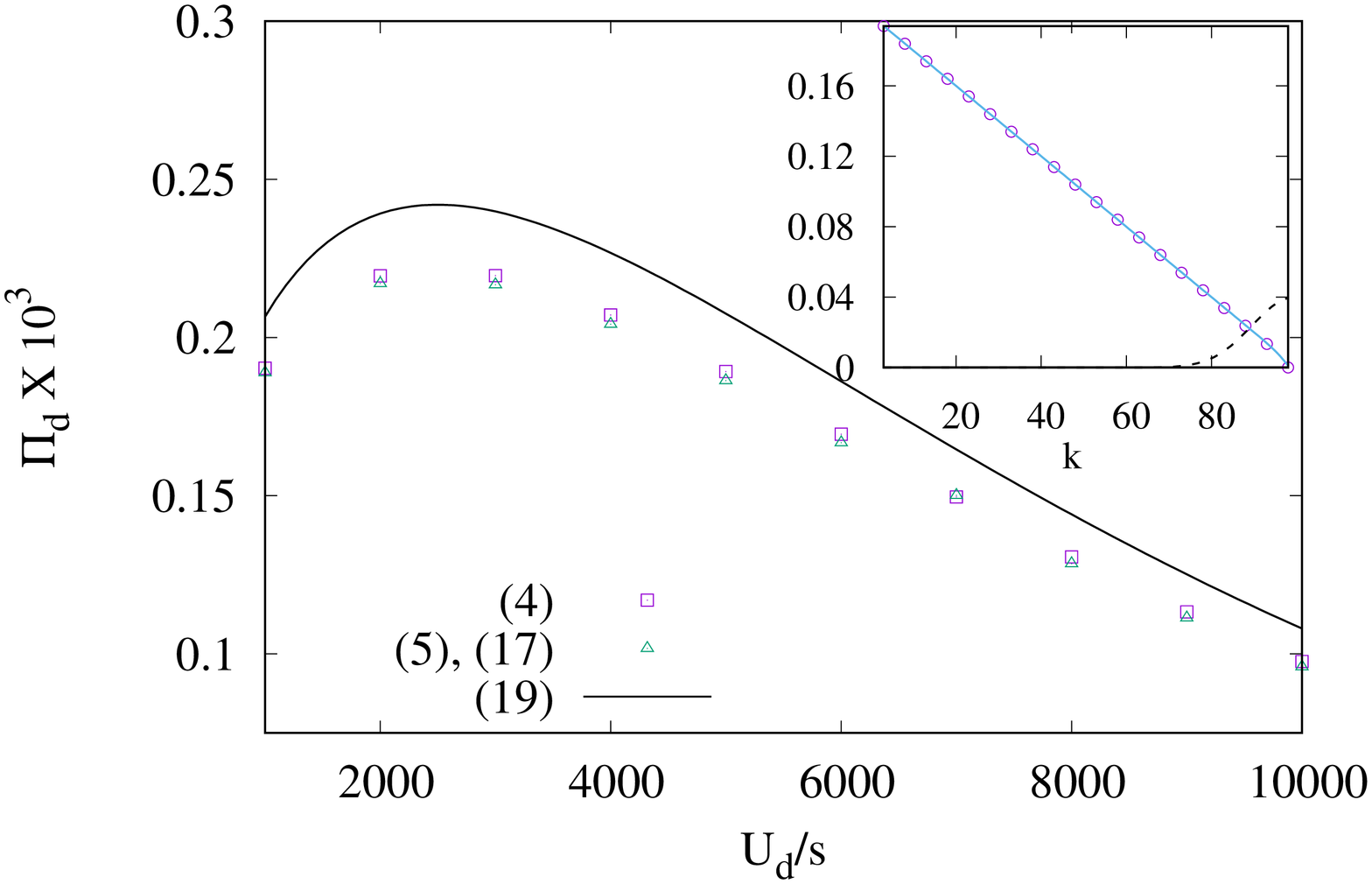}
\end{center}
\caption{Main: Nonmonotonic dependence of the total fixation probability $\Pi_d$ on mutator mutation rate $U_d$ for a fixed selection coefficient ($s=10^{-5}$) and mutator strength ($\lambda=50$). Inset: Variation of the fixation probability $\pi_d(k)$ with fitness class $k$. The data are obtained using (\ref{extprob2}) (points) and (\ref{case31}) (solid line) for $s=10^{-3}, U_d=0.1$. The \tc{black}{mutator frequency distribution $p_d(k)$} {(broken line)} is also shown.}
\label{fig_Qkprobs}
\end{figure}

\begin{figure} 
\begin{center}  
\includegraphics[width=0.8\textwidth,angle=270,keepaspectratio=true]{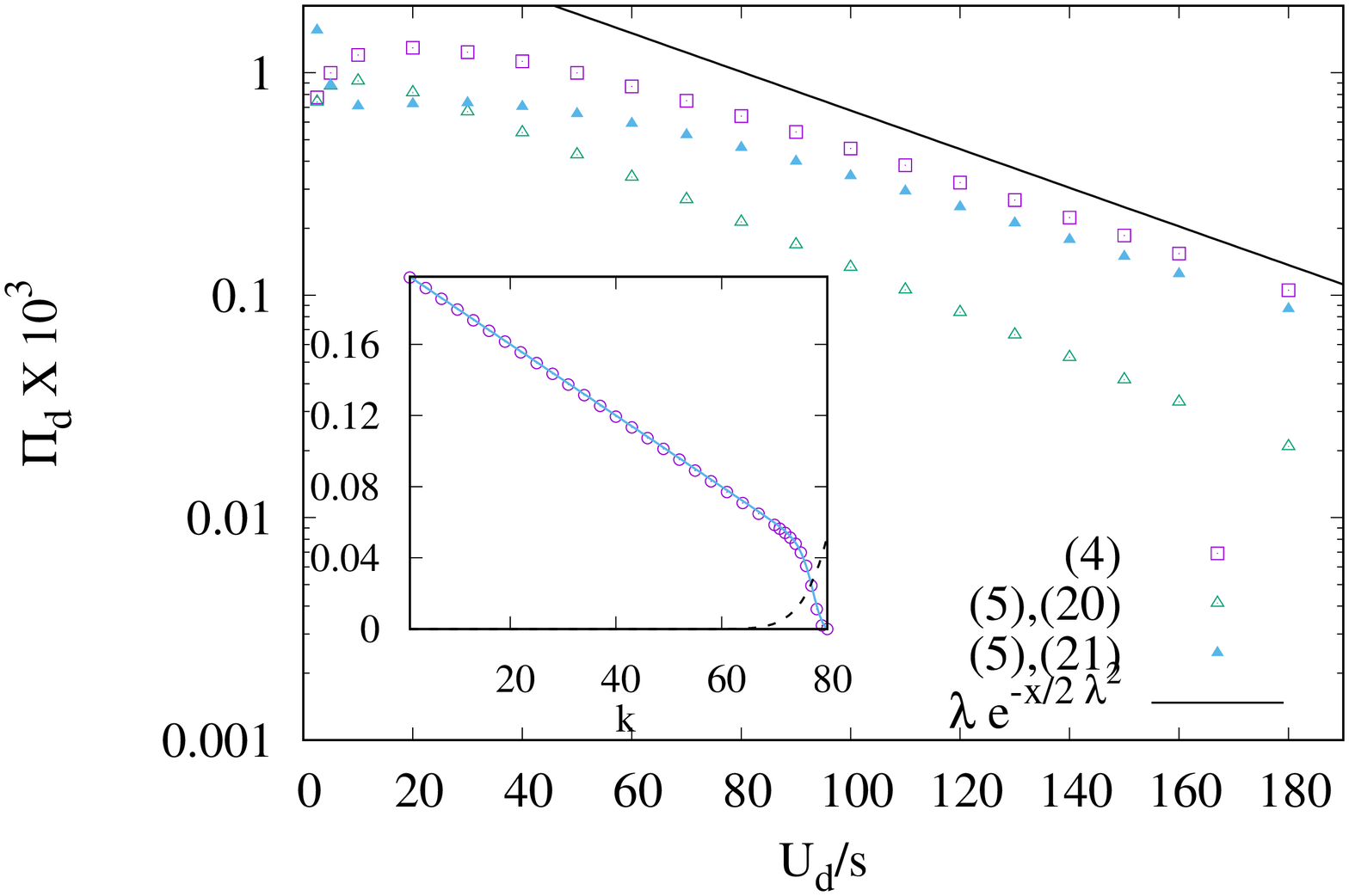}
\end{center}
\caption{Main: Nonmonotonic dependence of the total fixation probability $\Pi_d$ on mutator mutation rate $U_d$ for a fixed selection coefficient ($s=10^{-3}$) and mutator strength ($\lambda=5$). Inset: Variation of the fixation probability $\pi_d(k)$ with fitness class $k$. The data are obtained using (\ref{extprob2}) (points) and (\ref{case31}) (solid line) for $s=10^{-3}, U_d=0.1$. The \tc{black}{mutator frequency distribution $p_d(k)$} {(broken line)} is scaled by a factor $10$.}
\label{fig_Qkprobw}
\end{figure}

We again consider (\ref{scaled}) as both $s/U_d$ and $u_d/U_d$ are small here. Although $s/U_d$ is the smallest parameter, we can not neglect the term $s k/U_d$ in (\ref{scaled}) as it increases with the fitness class $k$.  To tackle this case, as described in Appendix~\ref{pi_nonlin}, we first obtain an approximate solution for $Q_d(k)$ for large fitness classes and then use its properties to arrive at an approximate solution for all the fitness classes which is given by 
\be
Q_d(k) = \frac{\pi_d (k)}{2 U_d} \approx 1 -\frac{s k}{U_d} - \frac{u_d}{U_d}  \left[1-\left(\frac{u_d}{s}\right)^{-2^{k+1-k_{\max}}} \right] ~.
\label{case31}
\ee
In the above expression, when $k \ll k_{\max}$, the last term on the RHS can be neglected and the fixation probability decreases linearly with the fitness class as in the last subsection. But for $k \lesssim k_{\max}$, the decay is faster than an exponential. 
Equation (\ref{case31}) for $\pi_d(k)$ is plotted in the inset of Figs.~\ref{fig_Qkprobs} and ~\ref{fig_Qkprobw} for strong and weak mutator, respectively, against the results obtained by solving (\ref{extprob2}) numerically, and we find a very good agreement. 

When the mutator is strong ($U_d \gg u_d$), the term containing $u_d/U_d$ on the RHS of (\ref{case31}) can be neglected and we find that the fixation probability decays linearly with the fitness class as indeed supported by the inset in Fig.~\ref{fig_Qkprobs}. Then, as shown in Appendix~\ref{pi_nonlin}, the total fixation probability (\ref{fptot}) is given by
\bea
\Pi_d & =& 2 \sum_{k=0}^{k_{\max}}  (U_d- s k) p_d(k) \\
&\approx& \sqrt{\frac{2 s U_d}{\pi}} e^{-\frac{U_d}{2 s \lambda^2}} ~.\label{twocases_s}
  \eea
When $U_d \ll 2 s \lambda^2$ or equivalently, $U_d \gg u_d^2/(2 s)$, the above equation shows that the fixation probability increases as $\sqrt{U_d}$ (as in the last subsection where $u_d < s < U_d$).  However, in the opposite parameter regime, $\Pi_d$  decreases exponentially with $U_d$. This can be understood as follows: Because of (\ref{popfrac_poisson}) for the \tc{black}{mutator frequency distribution $p_d(k)$}, the nonmutator is most likely to arise in fitness classes with $U_d/s - \sqrt{U_d/s} < k < U_d/s$ mutations. However, the fixation probability $\pi_d(k)$ is zero in this interval if $k_{\max} <  (U_d/s)-\sqrt{U_d/s}$ which, on using $k_{\max}\approx (U_d-u_d)/s$, implies that when $U_d >  s \lambda^2$, the chances of nonmutator fixation are considerably reduced.  
 Figure~\ref{fig_Qkprobs} shows that  expression (\ref{twocases_s}) matches well with the exact numerical calculations up to an additive constant as we have neglected the contribution from the $u_d$-dependent terms in (\ref{case31}). 
 
When the mutator is weak ($U_d \gtrsim u_d$), as the inset of Fig.~\ref{fig_Qkprobw} shows, the nonlinear decay of the fixation probability for large fitness classes can not be ignored. In Appendix~\ref{pi_nonlin}, the total fixation probability is calculated for  large $k_{\max}$ and we find that $\Pi_d=\Sigma_1+\Sigma_2$, where 
\bea
\Sigma_1 & =& 2 \sum_{k=0}^{k_{\max}} (U_d -s k - u_d) p_d(k) \propto  U_d^{-1/2} e^{-\frac{U_d}{2 s \lambda^2}} ~,\\
\Sigma_2 &=&  2 \sum_{k=0}^{k_{\max}} u_d \left(\frac{u_d}{s}\right)^{-2^{k+1-k_{\max}}}  p_d(k) \propto U_d^{1/2} e^{-\frac{U_d}{2 s \lambda^2}} \label{twocases_w2}~, 
\eea
and the proportionality constants are function of $\lambda$. We thus find that $\Pi_d \sim e^{-\frac{U_d}{2 s \lambda^2}}$ for $U_d/s \gg 2 \lambda^2$ in agreement with the data in Fig.~\ref{fig_Qkprobw}.

\subsection{Both deleterious and beneficial mutations}
\label{comp_mut}

We now consider the case when the beneficial mutation rates for mutator and nonmutator denoted by $U_b$ and $u_b$, respectively, are nonzero. It was first observed in  \citep{James:2016} that beneficial mutations in the mutator can {\it increase} the fixation probability of a nonmutator. This counterintuitive effect was shown in a limited parameter regime where $U_b \ll s$ and $u_b=0$. Our purpose here is to explore \tc{black}{the validity of} this result for a broader set of parameters. 

\subsubsection{\tc{black}{When mutation rates $u_b, u_d$ are nonzero}}

When both deleterious and beneficial mutations occur in the nonmutator, the mutation rate  $M(k \to j)$ from fitness class $k$ to $j$ in the nonmutator is given by 
\be
{M(k \to k+i) =} 
\begin{cases}
u_d \delta_{i,1}+ (1-u_d) \delta_{i,0} ~,~k=0 \\
u_d \delta_{i,1}+u_b \delta_{i,-1}+(1-u_b-u_d) \delta_{i,0} ~,~ k > 0~.
\end{cases}
\label{M2}
\ee
Using this in (\ref{extprob2}) for the fixation probability, we obtain 
\be
\label{fpk_wcomp1}
{1 - \pi(k) =}
\begin{cases}
\exp\left[-\frac{W(0)}{\overline W} \left(  (1-u_d) \pi(0) + u_d \pi(1) \right) \right]~,~ k=0\\
\exp \left[-\frac{W(k)}{\overline W} \left( u_b \pi(k-1) + (1-u_d-u_b) \pi(k) + u_d \pi(k+1) \right) \right] ~,~ k > 0~,
\end{cases}
\ee
where the average fitness of the mutator population  ${\overline W}=e^{-s {\bar k}}$ with ${\bar k}$ being the average number of deleterious mutations.  For  nonzero $U_b$, the mean ${\bar k}$  given by (\ref{meanexact}) is smaller than $U_d/s$, as one would intuitively expect. 

When beneficial mutations are ignored, by virtue of $u_b=0$, equation (\ref{fpk_wcomp1}) for the probability $\pi(k_{\max})$ closes ({\it i.e.}, it does not involve the fixation probability in any other fitness class) and therefore $\pi(k_{\max})$ can be determined.  This result then allows one to numerically calculate the fixation probability in lower fitness classes ($k < k_{\max}$) \citep{Johnson:2002}. However, for nonzero $u_b$, equation (\ref{fpk_wcomp1}) shows that for  $k > 0$, the probability $\pi(k)$ is coupled to the fixation probability in both the neighboring fitness classes. Thus if the fixation probability is zero beyond a fitness class $k_{\max}$, a calculation of $\pi(k_{\max})$ requires the knowledge of  $\pi(k_{\max}-1)$. For this reason, it is difficult to analyse (\ref{fpk_wcomp1}) even numerically and we have not been able to come up with an efficient method to do so. 

\begin{figure} 
\begin{center}  
\includegraphics[width=1\textwidth,angle=0,keepaspectratio=true]{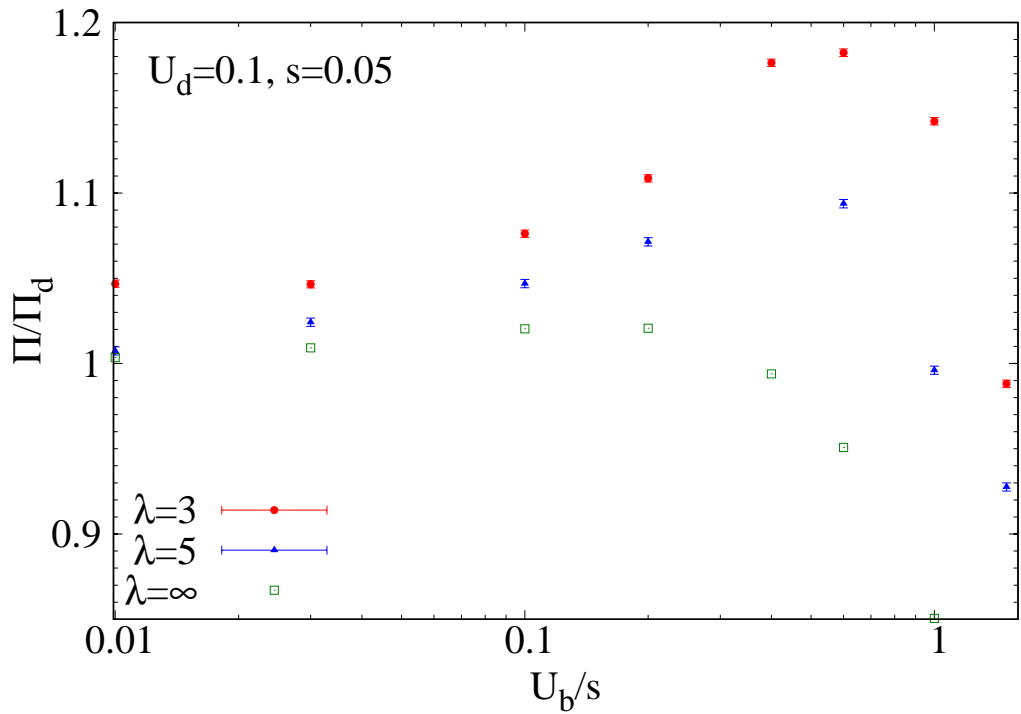}        
 \end{center}
\caption{Variation of the total fixation probability $\Pi$ when both beneficial and deleterious mutations are allowed relative to  $\Pi_d$ when only deleterious mutations occur as a function of the scaled mutation rate $U_b/s$.  For $\lambda=3$ and $5$, the simulation data for $N=5000$  and \tc{black}{for $\lambda \to \infty (u_d=0)$}, the numerical results obtained using (\ref{extprob2}) and (\ref{comp1}) are shown.}
\label{fig_benmut}
\end{figure}

However, using the simulation method described in Sec.~\ref{sto_sim}, we have studied this case for $U_d/s > 1$ and our results are shown in Fig.~\ref{fig_benmut}. We find that for large enough $U_b/s$, the total fixation probability $\Pi$ is smaller than $\Pi_d$; this behavior is expected as the mutational load carried by the mutators is reduced due to beneficial mutations resulting in a decrease in the fixation probability of nonmutator.  However, when $U_b/s$ is small, the fixation probability $\Pi > \Pi_d$ which is surprising. Figure~\ref{fig_benmut} also shows that the ratio $\Pi/\Pi_d$ is larger for weaker mutators and thus a lower bound on the ratio is obtained when $\lambda \to \infty$ (or, equivalently $u_b=u_d=0$). 

\subsubsection{\tc{black}{When mutation rates $u_b, u_d$ are zero}}

\tc{black}{For $u_d=u_b=0$,} as the surprising effect of beneficial mutations described above survives  \citep{James:2016} and the recursion equations for the probability $\pi(k)$ are amenable to analysis (see below), we now study this case using the multitype branching process.  

Assuming that the mutation rates and the selection coefficient are small, on proceeding in a manner similar to that in the last subsection, (\ref{fpk_wcomp1})  yields 
\be
\frac{\pi^2(k)}{2} \approx s ({\bar k}-  k) \pi(k) ~,~ k \geq 0~.
\ee
The nontrivial solution of the above equation is given by 
\be
\pi(k) = 2 s ( {\bar k}-  k)~,~ 0 \leq k  \leq  \lfloor \bar{k} \rfloor~.
\ee
The total fixation probability is then obtained as 
\be
\Pi= 2 s \sum_ {k=0}^{\lfloor \bar k \rfloor} ({\bar k}-  k) p(k)~,
\label{comp1}
\ee
where $p(k)$ is given by (\ref{popfrac_bessel}). In \citep{James:2016}, the frequency $p(k)$ was obtained by  solving (\ref{app_detmodel}) numerically; however, an exact expression for $p(k)$ that was obtained recently \citep{Jain:2016} allows us to extend our previous results (see (\ref{ben_new}) below). 

When $U_d/s \ll 1$, the average number of deleterious mutations is  smaller than one ($\lfloor \bar k \rfloor=0$) and we have $\Pi \approx 2 s  {\bar k}$ which, on using (\ref{meanapprox}), yields the relative fixation probability $\Pi/\Pi_d \approx 1- (U_b/s)$, in agreement with (11) of \citep{James:2016}. 

\begin{figure} 
\begin{center}  
\includegraphics[width=0.8\textwidth,angle=270,keepaspectratio=true]{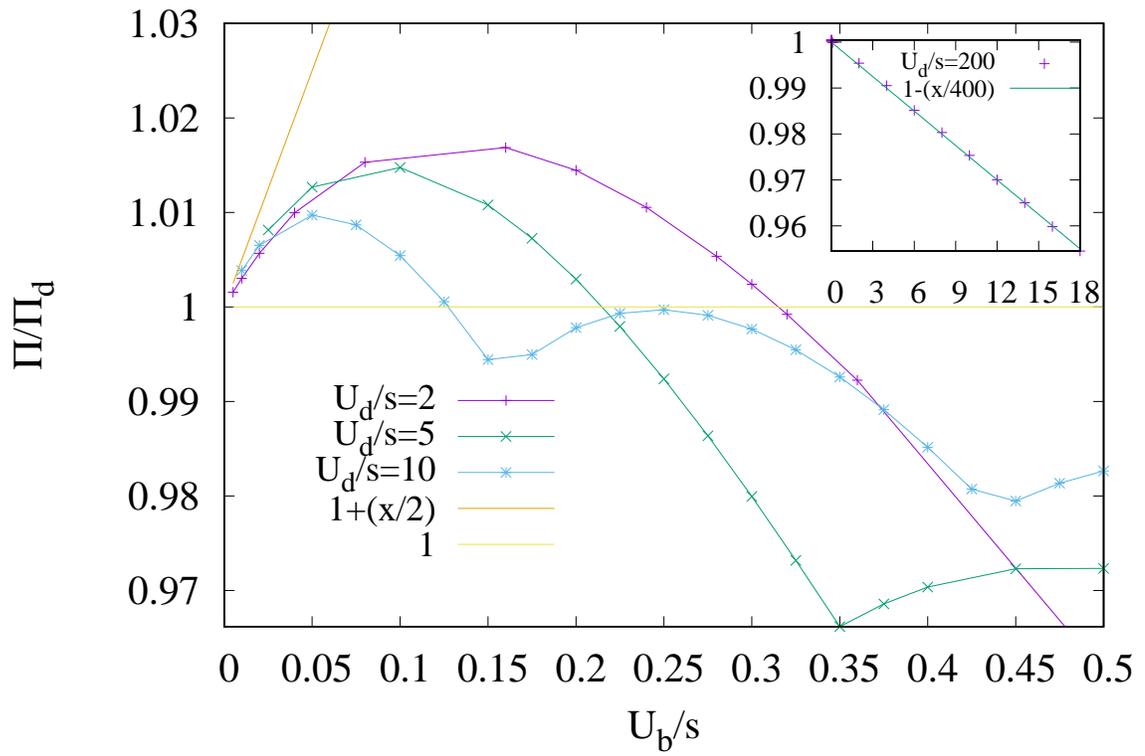}        
 \end{center}
\caption{Relative fixation probability $\Pi/\Pi_d$ obtained using (\ref{extprob2}) and (\ref{comp1}) when $U_d/s > 1$ and the mutation rates $u_b, u_d=0$. The inset also shows (\ref{ben_new}) for comparison.}
\label{fig_benmut2}
\end{figure}

When $U_d/s \gg 1$,  \tc{black}{we need to consider the parameter regimes $U_b/s \ll 1$ and $\gg 1$ separately. In the former case, the fixation probability $\Pi$ is larger than $\Pi_d$ as supported by a perturbation theory} in $U_b/s$ which yields $\Pi/\Pi_d =1+(U_b/s)$ \citep{James:2016}. But as Fig.~\ref{fig_benmut2} shows, this quantitative dependence agrees with the numerical calculations of (\ref{comp1}) in a narrow parameter range ($U_b/s \lesssim 0.01$) \tc{black}{while}  the relative fixation probability stays above one for $U_b/s < (U_d/s)^{-1}$. For larger $U_b/s$, however, the relative fixation probability is below unity. For $U_d/s \gg 1, U_b/s \gg 1$, the \tc{black}{mutator frequency distribution} is well approximated by a Gaussian function \citep{Jain:2016} with mean ${\bar k}$ and variance $\sigma^2$ given by (\ref{meanexact}) and (\ref{varexact}), respectively. Using these results in  (\ref{comp1}), we find that 
\be
\frac{\Pi}{\Pi_d} =\sqrt{\frac{\sigma^2}{U_d/s}} \approx 1- \frac{U_b}{2 U_d}~,
\label{ben_new}
\ee
which matches well with the numerical data  shown in the inset of Fig.~\ref{fig_benmut2}. The nonmonotonic change in $\Pi/\Pi_d$ is due to the discreteness of $k_{\max}$, as explained in Appendix~\ref{app_smallmax}. 


\section{Discussion}
\label{disc}

In this article, we have studied how the fixation probability of a rare nonmutator that arises in a large adapted population of asexual mutators depends on the mutation rates and the selection coefficient.  
Motivated by a long-term experiment on {\it E. coli} in which two nonmutator lineages arose in the mutator population when the fitness growth had slowed down considerably \citep{Wielgoss:2013}, we have calculated the fixation probability in the stationary state of the mutator population. 

Figure~\ref{fig_sdepn} shows that the fixation probability $\Pi_d$ (that ignores beneficial mutations) increases with selection coefficient $s$. This trend is consistent with the intuitive expectation that when the selective cost of a deleterious mutation is high, the mutation rate should be low. However, for sufficiently large $s$ where all mutations can be treated as lethal \citep{Lynch:2011}, the probability $\Pi_d$ saturates to a maximum value (see Sec.~\ref{reg1}).  
In this parameter regime, only one type of nonmutator - the one without any deleterious mutations - has a nonzero chance of fixation and the classic single-locus theory applies \citep{Haldane:1927b,Lynch:2011}. But for smaller selection coefficients, a nonmutator can carry many deleterious mutations and a multilocus analysis is required as has been done here using a multitype branching process \citep{Patwa:2008}. It is known that the fixation of a beneficial mutant is impeded due to interference from deleterious mutations when the loci are tightly linked \citep{Johnson:2002}. Viewing the nonmutator allele as a beneficial mutant under indirect selection, the reduction in the fixation probability in the multiloci scenario follows from this result.  The above discussion is also qualitatively consistent with the expectation that sexual populations in which loci are weakly linked are more likely to have low mutation rates \citep{Johnson:1999a,Tenaillon:2000,Raynes:2011}. 

As seen in Fig.~\ref{fig_lambda}, the fixation probability of the nonmutator also increases  with the mutator strength $\lambda=U_d/u_d$ where $U_d (u_d)$ is the deleterious mutation rate of the mutator (nonmutator); this is because a population of strong \tc{black}{mutators} would carry more deleterious mutations and hence more likely to get lost. To put it differently,  
a nonmutator with higher mutation rate has a lower chance of fixation because it also accrues deleterious mutations which weaken its advantage over the mutators. These expectations are indeed borne out by our analyses in Secs.~\ref{reg2} and \ref{reg3}. 

The above discussion suggests that a nonmutator with a mutation rate comparable to that of the mutator \citep{Mcdonald:2012,Wielgoss:2013} is  unlikely to fix.  However, from our detailed analysis in Sec.~\ref{reg3}, we arrive at a novel and important conclusion that a nonmutator is most likely to fix when the mutator population has a mutation rate $U_d \sim  s \lambda^2$ (also, see Figs.~\ref{fig_Qkprobs} and \ref{fig_Qkprobw}). \tc{black}{In a long term evolution experiment \citep{Wielgoss:2013}, the mutation rate of the ${\it E. coli}$ population carrying mutator allele decreased merely by a factor two at large times. As this event occurred in two independent lineages, using the data for $U_d$ and $\lambda$ given in Table 2 of \citep{Wielgoss:2013} in the above criteria for most probable mutation rate, we find the selection coefficient $s \sim 10^{-2}$ so that the selection to deleterious mutations is moderate in this experiment. We also note that for a wide range of selection coefficients $s=10^{-4}-10^{-2}$ and $U_d=0.03$ \citep{Wielgoss:2013}, the mutation rate is most likely to decrease by a factor $20$ or less; the above discussion thus provides an explanation for a small decrease in the mutation rates.}
The nonmonotonic behavior of the fixation probability with mutator's mutation rate, shown in Figs.~\ref{fig_Qkprobs} and \ref{fig_Qkprobw}, may be understood as follows: if $U_d$ is too small, the nonmutator does not offer significant advantage and therefore has a small chance of fixation. On the other hand, if  $U_d$ is too large, most individuals in the mutator population carry a large number of deleterious mutations. In this case, a nonmutator is favored only if it arises in mutator backgrounds with few deleterious mutations. But such mutator subpopulations are rare when $U_d$ is large thereby rendering the nonmutator fixation  unlikely (also, see the discussion below (\ref{twocases_s})). 

A factor studied here that can decrease the chances of nonmutator fixation is the occurrence of beneficial mutations in the mutator population as a result of which it carries lesser mutational load. However, we point out that when the mutator is weak, the beneficial mutations can enhance the fixation probability, see Fig.~\ref{fig_benmut}.  
As explained in \citep{James:2016}, this counterintuitive effect arises because of the competition between two opposing factors: while beneficial mutations  improve the fitness of the mutator population and hence adversely affect the fixation probability of the nonmutator, they also present a fitter mutator background for the nonmutator to arise thus augmenting its chance of fixation. This factor may also contribute to the fixation of the nonmutator in the experiment of  \citep{Wielgoss:2013} but  the relevant beneficial mutation rates do not seem to be available. 

Throughout this discussion, we have assumed that the mutator population is large enough so that random genetic drift may be ignored. A negative correlation between effective population size and  mutation rates has been found in recent studies \citep{Sung:2012,Lynch:2010b} and rationalised using a simple argument \citep{Lynch:2011,James:2016}; however, a more detailed and general analysis is needed for a better understanding of this relationship. 
For example, all the studies mentioned here except \citep{James:2016b} have examined the process of mutation rate reduction on a single-peaked, multiplicative fitness landscape; more complex fitness landscapes \citep{Visser:2014} could be used to understand the evolution of mutation rates. Moreover, the physiological costs \citep{Kimura:1967, Kondrashov:1995, Dawson:1998, Johnson:1999a, Baer:2007} associated with the decline in mutation rates can be included in a future  study to gain an insight into its effect on the mutation rate evolution.

\section*{Acknowledgements}
The authors thank Sona John for discussions during the early stages of this work. A. James thanks CSIR for the funding.

\clearpage
\appendix

\setcounter{section}{0}
\setcounter{equation}{0}
\makeatletter 
\renewcommand{\thesection}{A\@arabic\c@section} 
\renewcommand{\theequation}{A\@arabic\c@equation} 
\bigskip

\section{Deterministic model for \tc{black}{frequency distribution of mutators}}
\label{app_det}

For small selection coefficients, we can study the deterministic evolution of the \tc{black}{mutator frequency distribution $p(k,t)$} in continuous time \citep{Burger:2000}. The fitness $W(k)$ defined in the discrete time model (see (\ref{fitfunction})) and the fitness $w(k)$ in the continuous time model (discussed below) are related through $W(k)=e^{w(k)}$ which gives $w(k) \approx -s k$. Then the \tc{black}{mutator frequency distribution} obeys the following deterministic equations \citep{Jain:2016}, 
\bea
\frac{d p(0, t)}{dt} &=& U_b p(1, t) - U_d p(0, t) + s {\bar k}(t) p(0, t) ~,\\ 
\frac{d p(k, t)}{dt} &=&U_b p({k+1}, t) + U_d p(k-1, t) - (U_d+U_b) p(k, t) \nonumber \\
&& -s (k -{\bar k}(t))p(k, t)~,~k \geq 1~,
\label{app_detmodel}
\eea
where $\bar k=\sum_{k=0}^\infty k p(k, t)$ is the average number of deleterious mutations in the mutator population. The last term on the RHS of the above equations is the selection term while the other terms correspond to mutations in the neighboring fitness classes. In the stationary state where the LHS is zero, we will denote the frequency by $p(k)$. 

In the absence of beneficial mutations ($U_b=0$), the above equations simplify to yield the Poisson-distributed \tc{black}{mutator frequency} with mean $U_d/s$ \citep{Kimura:1966,Haigh:1978}:
\be
\label{popfrac_poisson}
p_d(k) = e^{-U_d/s} ~ \frac{(U_d/s)^k}{k!} ~.~
\ee
Thus the average fitness ${\overline W}_d= e^{-s \sum_{k=0}^\infty k p_d(k)} = e^{-U_d}$. 

When the beneficial mutations also occur, the stationary state population fraction is given by  \citep{Jain:2016} 
\be 
\label{popfrac_bessel}
p(k) = \frac{ (U_d/U_b)^{k/2} ~~ J_{k+\frac{(2-\zeta_0) \sqrt{U_b U_d}}{s}} \left(\frac{2 \sqrt{U_b U_d}}{s} \right)  }{ \sum_{m=0}^{\infty}  (U_d/U_b)^{m/2} ~~ J_{m+\frac{(2-\zeta_0) \sqrt{U_b U_d}}{s}} \left(\frac{2 \sqrt{U_b U_d}}{s} \right)  } ~.~
\ee
Here, $J_n(z)$ is the Bessel function of first kind with order $n$ and argument $z$ \citep{Abramowitz:1964} and $\zeta_0$ is the smallest root of the following equation:
\be
\label{lambda0}
J_{1+\frac{(2-\zeta) \sqrt{U_b U_d}}{s}} \left(\frac{2 \sqrt{U_b U_d}}{s} \right) - \left(2-\sqrt{\frac{U_b}{U_d}} - \zeta \right) ~~  J_{\frac{(2-\zeta) \sqrt{U_b U_d}}{s}} \left(\frac{2 \sqrt{U_b U_d}}{s} \right) = 0 ~.~ 
\ee
For $U_b=0$, the result (\ref{popfrac_bessel}) above reduces to (\ref{popfrac_poisson}) \citep{Jain:2016}. 

For completeness, below we quote the results obtained in \citep{Jain:2016} that are pertinent to the discussion here. When beneficial mutations are allowed, as one would expect, the average number of deleterious mutations \tc{black}{is} smaller than $U_d/s$ and given by  
\bea
{\bar k} &=& \frac{U_d+U_b}{s}- (2-\zeta_0) \frac{\sqrt{U_d U_b}}{s} ~,
\label{meanexact} \\
&\approx& \begin{cases}
\frac{U_d}{s} \left(1-\frac{U_b}{s} \right) ~,~ s \gg \sqrt{U_d U_b} \\
\frac{U_d-U_b}{s}-\left[\frac{2 \sqrt{U_b U_d}}{s} -\left(\frac{9 \pi}{8} \right)^{2/3} \left(\frac{\sqrt{U_b U_d}}{s}\right)^{1/3} + \frac{1}{1-\sqrt{U_b/U_d}}\right] ~,~ s \ll \sqrt{U_d U_b}~.
\label{meanapprox}
\end{cases}
\eea
For nonzero $U_b$, the variance of the number of deleterious mutations is larger than the mean and given by
\bea
\sigma^2 &=& \frac{U_d}{s}-\frac{U_b}{s} (1-p(0)) ~,
\label{varexact} \\
&\approx&\begin{cases}
\frac{U_d}{s} -\frac{U_b}{s} (1-e^{-U_d/s}) ~,~ s \gg \sqrt{U_d U_b} \\
\frac{U_d-U_b}{s} ~,~ s \ll \sqrt{U_d U_b} ~.
\label{varapprox}
\end{cases}
\eea

\section{When the integer $k_{\max}$ is small}
\label{app_smallmax}

Here we consider some simple cases where the fixation probability can be found exactly using the recursion equation  (\ref{fpk_weak_ws4}).

\noindent{$\underline{k_{\max}=0}$:} In this case, the difference in the mutation rates lie in the range $0 < U_d -u_d < s$. As $\pi_d(1)=0$, we immediately have $\pi_d(0)=2 (U_d-u_d)$ and therefore
\bea
\frac{\Pi_d}{s}&=&2 e^{-\mu} (\mu-\nu)~, \\
&=& 2 \mu e^{-\mu}  \left(1-\lambda^{-1} \right) ~,
\label{kmax0}
\eea
where, for brevity, we have defined 
\bea
\mu = U_d/s~,~ \nu = u_d/s~.
\label{munu}
\eea
From the above expression, we find that for a given $\lambda$, the total fixation probability first increases and then decreases as a function of $\mu$ when 
$0 < \mu < \lambda/(\lambda-1)$.
 
\noindent{$\underline{k_{\max}=1}$:} Using $\pi_d(2)=0$ in (\ref{fpk_weak_ws4}), we find the nonnegative solutions for the fixation probability to be
\bea
\pi_d(1) &=& 2 (U_d-u_d-s) ~,\\
\pi_d(0) &=& U_d-u_d+\sqrt{(U_d-u_d)^2+2 u_d \pi_d(1)}~.
\eea
As a result, the total fixation probability is given by
\be
\frac{\Pi_d}{s}=e^{-\mu}  \left[ (\mu-\nu)+2 \mu (\mu-\nu-1) +\sqrt{(\mu-\nu)^2 +4 \nu (\mu-\nu-1)} \right]~,
\label{kmax1}
\ee
and is a nonmonotonic function of the ratio $\mu$ when $\lambda/(\lambda-1)< \mu < 2 \lambda/(\lambda-1)$. 

\begin{figure} 
\begin{center}  
\includegraphics[width=1\textwidth,angle=0,keepaspectratio=true]{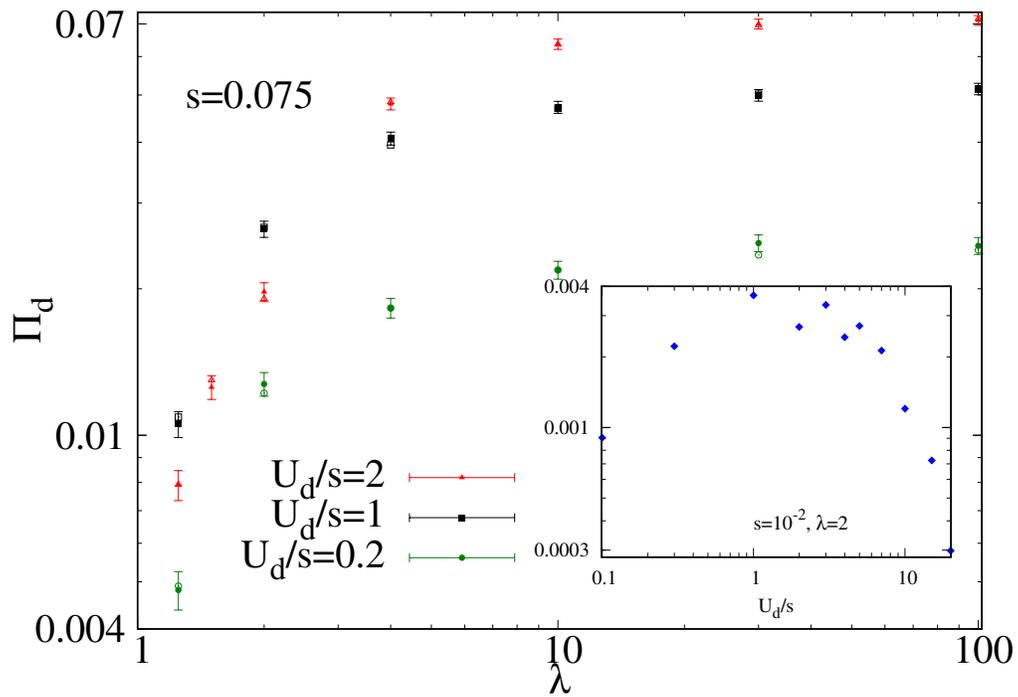}        
 \end{center}
\caption{Main: Variation of the total fixation probability with the mutator strength $\lambda$ when beneficial mutations are absent. The simulation data (\tc{black}{filled} symbols) for $N=5000$ and the numerical solution of (\ref{extprob2}) (\tc{black}{open} symbols) are shown. Inset: Nonmonotonic behavior of the fixation probability $\Pi_d$ as a function of the scaled mutation rate $U_d/s$ for $0 \leq k_{\max} \leq 10$.}
\label{fig_lambda}
\end{figure}

From the above simple examples and the inset of Fig.~\ref{fig_lambda}, we conclude that when $k_{\max}$ is small,  the total fixation probability increases and decreases several times as $U_d/s$ is varied. However, as Fig.~\ref{fig_lambda} also shows, this effect is unimportant for large $k_{\max}$. 


\section{Fixation probability when $u_d < s < U_d$}
\label{app_case2w}

Using (\ref{popfrac_poisson}) and (\ref{case2wk}) in the sum (\ref{fptot}), we find that the total fixation probability is given by 
\be
\Pi_d=2 \sum_{k=0}^{k_{max}} \left( U_d- s k - \frac{u_d s}{U_d- s k} \right) ~ p_d(k)~.
\label{app_pert2}
\ee
We first note that for $u_d=0$, the sum in the above equation can be done exactly \citep{James:2016} and is given by
\be
\Pi_d= \frac{e^{-U_d/s}}{k_{\max}!} \left(\frac{U_d}{s} \right)^{1+k_{\max}}~.
\label{exact0}
\ee
We verify that the results (\ref{kmax0}) and (\ref{kmax1}) for $k_{\max}=0, 1$, respectively, are reproduced from the above equation when $u_d=0$.

To estimate the fixation probability $\Pi_d$, we first approximate the Poisson distribution $p_d(k)$ with mean and variance $U_d/s$ by a Gaussian distribution with these properties \citep{James:2016}. On approximating the sum in 
(\ref{app_pert2}) by an integral, we get
\bea
\Pi_d &\approx& 2 s \int_{0}^{\mu-\nu} dk \left(\mu-k - \frac{\nu}{\mu-k} \right) 
~\frac{e^{-\frac{(k-\mu)^2}{2 \mu}}}{\sqrt{2 \pi \mu}} \\
&\approx& s \sqrt{\frac{2 \mu}{\pi}} \int_{\frac{\nu^2}{2 \mu}}^{\frac{\mu}{2}}  dz ~e^{-z} \left(1-\frac{\nu}{2 \mu z} \right) ~,\label{app_pert3}
\eea
where $\mu=U_d/s$ and $\nu=u_d/s$ as defined in (\ref{munu}). For small $\nu$ and large $\mu$,  on carrying out the above integrals, we obtain 
\bea
\Pi_d = s \sqrt{\frac{2 \mu}{\pi}} \left[1- \frac{\nu}{2 \mu} \ln \left(\frac{2 \mu}{\nu^2} \right) \right]~.
\label{app_case2f}
\eea
The second term in the bracket on the RHS is obtained on using that the exponential integral $E_1(z)=\int_z^\infty dt~e^{-t}/t \approx -\ln z$ for small $z$ (see 5.1.1 and 5.1.11, \citep{Abramowitz:1964}). We remark that although (\ref{app_pert2}) is linear in the small parameter $\nu$, the correction term is nonlinear due to the logarithmically-diverging second integral in (\ref{app_pert3}). 
\section{Fixation probability when $s < u_d < U_d$}
\label{pi_nonlin}

We first study the behavior of the fixation probability for fitness classes close to $k_{\max}$. Using $Q_d(k_{\max}+1)=0$ and $k_{\max} \approx (U_d-u_d)/s$ for large $k_{\max}$ (see the discussion after (\ref{fpk_weak_ws4})) in equation (\ref{scaled}), we get  $Q_d(k_{\max})=0$ which further yields $Q_d(k_{\max}-1)=s/U_d$. To obtain simple expressions for other fitness classes, we note that for fitness classes close to the boundary $k_{\max}$, the coefficient of $Q_d(k)$ in (\ref{scaled}) can be neglected yielding a simpler recursion equation, 
\be
Q^{>}_d(k) \approx \sqrt{Q^{>}_d(k+1)/\lambda} ~,
\label{app_recexpo}
\ee
which can be easily solved subject to the boundary condition at $k=k_{\max}-1$ and we get
\bea
Q^{>}_d(k) &=& \frac{1}{\lambda}~\left[\lambda Q_d(k_{\max}-1) \right]^{2^{k+1-k_{\max}}} \\
&=& \frac{1}{\lambda}~\left(\frac{s}{u_d}\right)^{2^{k+1-k_{\max}}} ~.
\label{app_gtr1}
\eea
Note that the above fixation probability is of a double exponential form ($e^{-e^{-x}}$) and therefore decays faster than an exponential as the \tc{black}{fitness class $k$ increases towards $k_{\max}$.} Furthermore, for large $k_{\max}-k$, it saturates to $\lambda^{-1}$. 

To find the behavior for small $k$, we now write the complete solution $Q_d(k)$ as 
\be
Q_d(k)=Q^<_d(k)+Q^>_d(k) ~.
\label{app_sum1}
\ee
Substituting this in (\ref{scaled}), we get
\bea
&&\left[Q^<_d(k)- \left(1-\frac{s k}{U_d}- \frac{u_d}{U_d} \right) \right]Q_d(k) \\
&=& \lambda^{-1} Q^<_d(k+1) - Q^<_d(k) Q^>_d(k) \label{app_ls1}\\
&\approx& \lambda^{-1} \left[Q^<_d(k+1) - Q^<_d(k) \right]  ~.
\label{app_ls2}
\eea
\tc{black}{The expression} (\ref{app_ls1}) is exact for $Q^<_d(k)$. But as \tc{black}{it is} nonlinear, we find an approximate solution for $Q^<_d(k)$ using the fact that $Q^>_d(k)$ saturates quickly to $\lambda^{-1}$  to arrive at (\ref{app_ls2}). If we now set (\ref{app_ls2}) to zero, we get
\be
Q^<_d(k)=1-\frac{s k}{U_d}- \frac{u_d}{U_d}~.
\label{app_less1}
\ee
The above solution shows that the error committed in ignoring (\ref{app_ls2}) is of the order $s u_d/U_d^2$ which is negligible  in the parameter regime under consideration. Using (\ref{app_gtr1}) and (\ref{app_less1}) in (\ref{app_sum1}), we finally obtain the fixation probability in (\ref{case31}). 

To find the total fixation probability, we need to perform the following sums, 
\be
\frac{\Pi_d}{2 s} =  \sum_{k=0}^{k_{max}} (k_{\max}-k) p_d(k)+ \nu \sum_{k=0}^{k_{max}-1}\nu^{-2^{k+1-k_{\max}}}  ~ p_d(k)  ~,
\label{app_Pid32}
\ee
where $p_d(k)$ is the Poisson distribution given by (\ref{popfrac_poisson}) and $\mu, \nu$ are defined in (\ref{munu}).  

\noindent{\it Strong mutator ($\lambda \gg 1$):} Neglecting the mutation rate of the nonmutator in (\ref{app_Pid32}) when $U_d \gg u_d$, we find that
\bea
\frac{\Pi_d}{2 s} &\approx& \sum_{k=0}^{k_{\max}} (\mu-k) \frac{e^{-\mu} \mu^k}{k!} ~, \label{junk}\\
&=&  \frac{e^{-\mu} \mu^{k_{\max}+1}}{k_{\max}!} ~, \\
&\approx& \sqrt{\frac{\mu}{2 \pi}} e^{-\frac{\mu}{2 \lambda^2}}~.
\label{app_gauss2}
\eea
Alternatively, in (\ref{junk}), we can approximate the Poisson distribution by a Gaussian and the sum by an integral (as in Appendix~\ref{app_case2w}) to get the above result.

\noindent{\it Weak mutator ($\lambda \gtrsim 1$):} The first sum on the RHS of the above equation can be expressed as ((26.4.21), \citep{Abramowitz:1964})
\bea
&& \sum_{k=0}^{k_{max}} (k_{\max}-k) p_d(k) \\
&=& \frac{e^{-\mu} \mu^{k_{\max}+1}}{k_{\max}!} \left[1- \nu \int_0^\infty dt ~(1+t)^{k_{\max}} ~e^{-\mu t} \right] ~,\\
&=& \frac{\nu e^{-\mu} \mu^{k_{\max}+1}}{k_{\max}!} \int_0^\infty dt ~e^{-\nu t} 
\left[1- (1+t)^{k_{\max}} e^{-k_{\max} t} \right]  \label{app_tvar} ~,\\
&\approx& \frac{\nu e^{-\mu} \mu^{k_{\max}+1}}{k_{\max}!} \int_0^\infty dt ~e^{-\nu t} 
\left[1- e^{-k_{\max} t^2/2} \right] ~,
\eea
where we have used that when $\nu \gg 1$, the integrand in (\ref{app_tvar}) is appreciable for $t \ll 1$. Solving the integral in the last equation, we obtain
\bea
&& \sum_{k=0}^{k_{max}} (k_{\max}-k) p_d(k) \\
&=&\frac{e^{-\mu} \mu^{k_{\max}+1}}{k_{\max}!} \left[ 1- \sqrt{\frac{\pi \nu^2}{2 k_{\max}}} e^{\frac{\nu^2}{2 k_{\max}}} \textrm{erfc}\left(\frac{\nu}{\sqrt{2 k_{\max}}} \right) \right] ~, \\
&=&\begin{cases}
 \frac{\mu}{\sqrt{2 \pi k_{\max}}}~,~ 1 \ll \mu \ll 2 \lambda^2 \\
 \frac{\mu e^{-\frac{\mu}{2 \lambda^2}}}{\sqrt{2 \pi k_{\max}}} ~\frac{k_{\max}}{\nu^2} ~,~ \mu \gg 2 \lambda^2 ~,
\end{cases}
 \eea
where $\textrm{erfc}(x)=(2/\sqrt{\pi}) \int_x^\infty dt ~e^{-t^2}$ is the complementary error function and we have used the asymptotic expansion of error function to obtain the last expression ((7.1.23), \citep{Abramowitz:1964}). 
 
The second sum on the RHS of (\ref{app_Pid32}) can be estimated via a saddle-point method  \citep{Arfken:1985}. Denoting the logarithm of the summand by $f(k)$ and expanding it about its turning point at $k_*$ where $f'(k_*)=0$, we have 
\bea
&&\sum_{k=0}^{k_{max}-1} \nu^{-2^{k+1-k_{\max}}}  ~ p_d(k) ~\\
&\approx& \int_{0}^{k_{max}} dk~ e^{-f(k_*)-\frac{(k-k_*)^2}{2} f''(k_*)} ~\\
&=& e^{-f(k_*)} \sqrt{\frac{\pi}{2 f''(k_*)}} \left[\textrm{erf}\left({\frac{k_*\sqrt{f''(k_*)}}{\sqrt{2}}} \right) + \textrm{erf}\left({\frac{(k_{\max}-k_*)\sqrt{f''(k_*)}}{\sqrt{2}}} \right)\right]~,
\label{app_saddle}
\eea
where prime denotes a derivative with respect to $k$. In the above equation, $k_*$ is a solution of the following equation, \bea
\ln \nu~ \ln 2 ~2^{k_*+1-k_{max}}=\ln(\mu/k_*)~,
\label{kstareqn}
\eea
and the second derivative is given by 
\be
f''(k_*)=\ln 2 \ln (\mu/k_*)+ k_*^{-1}~.
\ee
Taking logarithms on both sides of (\ref{kstareqn}), we find that 
\be 
k_* \approx k_{\max}-1 -(\ln(\ln \nu)/\ln 2) ~,
\label{app_kstar}
\ee
which shows that $k_*$ is close to $k_{\max}$. 
This result furthermore yields $f''(k_*) \sim \lambda^{-1}$. Since the validity of the saddle-point method requires $f''(k_*)$ to be large  \citep{Arfken:1985}, our approximation is good when $\lambda$ is not too large.  
We thus have
\bea
&& \nu \sum_{k=0}^{k_{max}-1} \nu^{-2^{k+1-k_{\max}}}  ~ p_d(k) \label{junk22}\\
&\approx& \frac{e^{-\mu} \mu^{k_*+1}}{k_*!} \left(\frac{k_*}{\mu} \right)^{1/\ln 2} \frac{\nu}{\mu} \sqrt{\frac{2 \pi}{\ln 2 \ln (\mu/k_*)}} ~,\\
&\approx&  \frac{\nu e^{-\mu} \mu^{k_{\max}}}{k_{\max}!} ~\sqrt{\frac{2 \pi \lambda}{\ln 2}} 
\left(\frac{k_{\max}}{\mu} \right)^{k_{\max}-k_*+(\ln 2)^{-1}} ~,\\
&\approx& \sqrt{\mu} e^{-\frac{\mu}{2 \lambda^2}} \left[\sqrt{\frac{1}{(\lambda-1) \ln 2}} 
\left(\frac{k_{\max}}{\mu} \right)^{k_{\max}-k_*+(\ln 2)^{-1}} \right]~. 
\eea
Note that in the last equation, the factors in the bracket  are a function of $\lambda$ and therefore the sum in (\ref{junk22}) is proportional to $\sqrt{\mu} e^{-\frac{\mu}{2 \lambda^2}}$ for fixed $\lambda$.

\clearpage

\clearpage

\clearpage

\clearpage

\clearpage

\clearpage

\clearpage





\begin{thebibliography}{10}

\bibitem{Abramowitz:1964}
M.~Abramowitz and I.~A. Stegun.
\newblock {\em Handbook of Mathematical Functions with Formulas, Graphs, and
  Mathematical Tables}.
\newblock Dover, 1964.

\bibitem{Arfken:1985}
G.~Arfken.
\newblock {\em Mathematical Methods for Physicists}.
\newblock Academic Press, New York, 1985.

\bibitem{Baer:2007}
C.~F. Baer, M.~M. Miyamoto, and D.~R. Denver.
\newblock Mutation rate variation in multicellular eukaryotes: Causes and
  consequences.
\newblock {\em Nat. Rev. Genet.}, 8:619--631, 2007.

\bibitem{Burger:2000}
R.~B{\"u}rger.
\newblock {\em The Mathematical Theory of Selection, Recombination, and
  Mutation}.
\newblock Wiley, Chichester, 2000.

\bibitem{Chao:1983}
L.~Chao and E.~C. Cox.
\newblock Competition between high and low mutating strains of {{\it
  Escherichia coli}}.
\newblock {\em Evolution}, 37:125--134, 1983.

\bibitem{Dawson:1998}
K.J. Dawson.
\newblock Evolutionarily stable mutation rates.
\newblock {\em J. theor. Biol.}, 194:143--157, 1998.

\bibitem{Visser:2014}
J.~A. G.~M. de~Visser and J.~Krug.
\newblock Empirical fitness landscapes and the predictability of evolution.
\newblock {\em Nat. Rev. Genet.}, 15:480--490, 2014.

\bibitem{Desai:2011}
M.M. Desai and D.S. Fisher.
\newblock The balance between mutators and nonmutators in asexual populations.
\newblock {\em Genetics}, 188:997--1014, 2011.

\bibitem{Ewens:1979}
W.J. Ewens.
\newblock {\em Mathematical Population Genetics}.
\newblock Springer, Berlin, 1979.

\bibitem{Giraud:2001}
A.~Giraud, I.~Matic, O.~Tenaillon, A.~Clara, M.~Radman, M.~Fons, and F.~Taddei.
\newblock Costs and benefits of high mutation rates: adaptive evolution of
  bacteria in the mouse gut.
\newblock {\em Science}, 291:2606 -- 2608, 2001.

\bibitem{Good:2016}
B.H. Good and M.~Desai.
\newblock Evolution of mutation rates in rapidly adapting asexual populations.
\newblock {\em Genetics}, 204:1249--1266, 2016.

\bibitem{Haigh:1978}
J.~Haigh.
\newblock The accumulation of deleterious genes in a population - {M}uller's
  ratchet.
\newblock {\em Theoret. Population Biol.}, 14:251--267, 1978.

\bibitem{Haldane:1927b}
J.~B.~S. Haldane.
\newblock A mathematical theory of natural and artificial selection. v.
\newblock {\em Proc. Camb. Philos. Soc.}, 23:838--844, 1927.

\bibitem{Harris:1963}
T.E. Harris.
\newblock {\em The theory of branching processes}.
\newblock Springer-Verlag Berlin Heidelberg, 1963.

\bibitem{Jain:2008b}
K.~Jain.
\newblock Loss of least-loaded class in asexual populations due to drift and
  epistasis.
\newblock {\em Genetics}, 179:2125, 2008.

\bibitem{Jain:2016}
K.~Jain and S.~John.
\newblock Deterministic evolution of an asexual population under the action of
  beneficial and deleterious mutations on additive fitness landscapes.
\newblock {\em Theo. Pop. Biol.}, 112:117--125, 2016.

\bibitem{Jain:2013}
K.~Jain and A.~Nagar.
\newblock Fixation of mutators in asexual populations: the role of genetic
  drift and epistasis.
\newblock {\em Evolution}, 67:1143--1154, 2013.

\bibitem{James:2016b}
A.~James.
\newblock Fixation probability of rare nonmutator and evolution of mutation
  rates.
\newblock {\em J. theor. Biol.}, 407:225--237, 2016.

\bibitem{James:2016}
A.~James and K.~Jain.
\newblock Fixation probability of rare nonmutator and evolution of mutation
  rates.
\newblock {\em Ecology and Evolution}, 6:755--764, 2016.

\bibitem{Johnson:1999a}
T.~Johnson.
\newblock Beneficial mutations, hitchhiking and the evolution of mutation rates
  in sexual populations.
\newblock {\em Genetics}, 51:1621--1631, 1999.

\bibitem{Johnson:2002}
T.~Johnson and N.H. Barton.
\newblock The effect of deleterious alleles on adaptation in asexual
  populations.
\newblock {\em Genetics}, 162:395--411, 2002.

\bibitem{Kimura:1967}
M.~Kimura.
\newblock On the evolutionary adjustment of spontaneous mutation rates.
\newblock {\em Genet. Res.}, 9:23--34, 1967.

\bibitem{Kimura:1966}
M.~Kimura and T.~Maruyama.
\newblock The mutational load with epistatic gene interactions in fitness.
\newblock {\em Genetics}, 54:1337--1351, 1966.

\bibitem{Kondrashov:1995}
A.~Kondrashov.
\newblock Contamination of the genome by very slightly deleterious mutations:
  why have we not died 100 times over?
\newblock {\em J. theor. Biol.}, 175:583--594, 1995.

\bibitem{Leigh:1973}
E.G. Leigh.
\newblock The evolution of mutation rates.
\newblock {\em Genetics}, 73:1--18, 1973.

\bibitem{Lynch:2010b}
M.~Lynch.
\newblock Evolution of the mutation rate.
\newblock {\em Trends in Genetics}, 26:345--352, 2010.

\bibitem{Lynch:2011}
M.~Lynch.
\newblock The lower bound to the evolution of mutation rates.
\newblock {\em Genome Evol. Biol.}, 3:1107--1118, 2011.

\bibitem{Smith:1974}
J.~Maynard~Smith and J.~Haigh.
\newblock Hitchhiking effect of a favourable gene.
\newblock {\em Genet. Res.}, 23:23--35, 1974.

\bibitem{Mcdonald:2012}
M.J. McDonald, Y.-Y. Hsieh, Y.-H. Yu, S.-L. Chang, and J.-Y. Leu.
\newblock The evolution of low mutation rates in experimental mutator
  populations of \emph{Saccharomyces cerevisiae}.
\newblock {\em Current Biology}, 22:1235--1240, 2012.

\bibitem{Notleymcrobb:2002}
L.~Notley-McRobb, S.~Seeto, and T.~Ferenci.
\newblock Enrichment and elimination of \emph{mutY} mutators in
  \emph{Escherichia coli} populations.
\newblock {\em Genetics}, 162:1055--1062, 2002.

\bibitem{Oliver:2000}
A.~Oliver, R.~Cant{\'o}n, P.~Campo, F.~Baquero, and J.~Bl{\'a}zquez.
\newblock High frequency of hypermutable \emph{Pseudomonas aeruginosa} in
  cystic fibrosis lung infection.
\newblock {\em Science}, 288:1251--1253, 2000.

\bibitem{Palmer:2006}
M.E Palmer and M.~Lipsitch.
\newblock The influence of hitchhiking and deleterious mutation upon asexual
  mutation rates.
\newblock {\em Genetics}, 173:461--472, 2006.

\bibitem{Patwa:2008}
Z.~Patwa and L.M. Wahl.
\newblock The fixation probability of beneficial mutations.
\newblock {\em Journal of the Royal Society Interface}, 5:1279--1289, 2008.

\bibitem{Perfeito:2007}
L.~Perfeito, L.~Fernandes, C.~Mota, and I.~Gordo.
\newblock Adaptive mutations in bacteria: high rate and small effects.
\newblock {\em Science}, 317:813--815, 2007.

\bibitem{Raynes:2011}
Y.~Raynes, M.R. Gazzara, and P.D. Sniegowski.
\newblock Mutator dynamics in sexual and asexual experimental populations of
  yeast.
\newblock {\em BMC Evolutionary Biology}, 11:158, 2011.

\bibitem{Raynes:2014}
Y.~Raynes and P.D. Sniegowski.
\newblock Experimental evolution and the dynamics of genomic mutation rate
  modifiers.
\newblock {\em Heredity}, 113:375--380, 2014.

\bibitem{Singh:2017}
T.~Singh, M.~Hyun, and P.~Sniegowski.
\newblock Evolution of mutation rates in hypermutable populations of
  \emph{Escherichia coli} propagated at very small effective population size.
\newblock {\em Biol. Lett.}, 13:20160849, 2017.

\bibitem{Sniegowski:2010}
P.~D. Sniegowski and P.~J. Gerrish.
\newblock Beneficial mutations and the dynamics of adaptation in asexual
  populations.
\newblock {\em Phil. Trans. R. Soc. B}, 365:1255--1263, 2010.

\bibitem{Sniegowski:1997}
P.~D. Sniegowski, P.~J. Gerrish, and R.E. Lenski.
\newblock Evolution of high mutation rates in experimental populations of
  \emph{ E. coli}.
\newblock {\em Nature}, 387:703--705, 1997.

\bibitem{Sturtevant:1937}
A.H. Sturtevant.
\newblock Essays on evolution i. on the effects of selection on the mutation
  rate.
\newblock {\em Q. Rev. Biol.}, 12:464--476, 1937.

\bibitem{Sung:2012}
W.~Sung, M.~S. Ackerman, S.~F. Miller, T.~G. Doak, and M.~Lynch.
\newblock The drift-barrier hypothesis and mutation-rate evolution.
\newblock {\em Proc. Natl. Acad. Sci. USA}, 109:18488--18492, 2012.

\bibitem{Taddei:1997}
T.~Taddei, M.~Radman, J.~Maynard-Smith, B.~Toupance, P.~H. Gouyon, and
  B.~Godelle.
\newblock Role of mutator alleles in adaptive evolution.
\newblock {\em Nature}, 387:700--702, 1997.

\bibitem{Tenaillon:2000}
O.~Tenaillon, H.~Le~Nagard, B.~Godelle, and F.~Taddei.
\newblock Mutators and sex in bacteria: Conflict between adaptive strategies.
\newblock {\em Proc. Natl. Acad. Sci. USA}, 152:485--493, 1999.

\bibitem{Tenaillon:1999}
O.~Tenaillon, B.~Toupance, H.L. Nagard, F.~Taddei, and B.~Godelle.
\newblock Mutators, population size, adaptive landscape and the adaptation of
  asexual populations of bacteria.
\newblock {\em Genetics}, 152:485--493, 1999.

\bibitem{Trobner:1984}
W.~Tr{\"o}bner and R.~Piechocki.
\newblock Selection against hypermutability in {{\it Escherichia coli}} during
  long-term evolution.
\newblock {\em Mol. Gen. Genet.}, 198:177--178, 1984.

\bibitem{Wielgoss:2013}
S.~Wielgoss, J.E. Barrick, O.~Tenaillon, M.J. Wiser, W.J. Dittmar,
  S.~Cruveiller, B.~Chane-Woon-Ming, C.~M{\'e}digue, R.~E. Lenski, and
  D.~Schneider.
\newblock Mutation rate dynamics in a bacterial population reflect tension
  between adaptation and genetic load.
\newblock {\em Proc. Natl. Acad. Sci USA}, 110:222--227, 2013.

\bibitem{Wylie:2009}
C.S. Wylie, C.-M. Ghim, D.~Kessler, and H.~Levine.
\newblock The fixation probability of rare mutators in finite asexual
  populations.
\newblock {\em Genetics}, 181:1595--1612, 2009.

\end{thebibliography}
\end{document}